\PassOptionsToPackage{pdfpagelabels=false}{hyperref}
\RequirePackage{fix-cm}
\documentclass[fleqn,usenatbib]{mnras}

\usepackage[T1]{fontenc}
\usepackage{ae,aecompl}

\usepackage{graphicx}	
\usepackage{amsmath}	
\usepackage{amssymb}	
\usepackage{epstopdf} 
\usepackage[para]{threeparttable} 
\usepackage{subfig}
\usepackage{pstricks}
\usepackage{float}

\newcommand{\pri}    {${\rlap.}^{\prime \prime}$}
\newcommand{\prp}    {${\rlap.}^{\prime}$}

\title[2008 OG$_{19}$: A highly elongated Trans-Neptunian Object]{2008 OG$_{19}$: A highly elongated Trans-Neptunian Object}

\author[E. Fern\'andez-Valenzuela]{
E. Fern\'andez-Valenzuela,$^{1}$\thanks{E-mail: estela@iaa.es (IAA)}
J. L. Ortiz,$^{1}$
R. Duffard,$^{1}$
P. Santos-Sanz,$^{1}$
N. Morales.$^{1}$
\\
$^{1}$Instituto de Astrof\'{i}sica de Andaluc\'{i}a, IAA-CSIC, Glorieta de la Astronom\'{i}a s/n, 18008 Granada, Spain
}

\date{Accepted XXX. Received YYY; in original form ZZZ}

\pubyear{2015}

\begin{document}
\label{firstpage}
\pagerange{\pageref{firstpage}--\pageref{lastpage}}
\maketitle

\begin{abstract}
From two observing runs during the 2014 summer at Calar Alto Observatory in Almer\'ia (Spain) and at Sierra Nevada Observatory in Granada (Spain), we were able to derive CCD photometry of the Trans-Neptunian Object 2008 OG$_{19}$. We analyzed the time series and obtained a double-peaked light curve with a peak to valley amplitude of (0.437 $\pm$ 0.011) mag and a rotational period of (8.727$\pm$ 0.003) h. This implies that this object is very elongated, closely resembling Varuna's case. The photometry also allowed us to obtain an absolute magnitude in R-band of (4.39 $\pm$ 0.07) mag. From this result we estimated an equivalent diameter of 2008 OG$_{19}$ which is 619$^{+56}_{-113}$ km using an average albedo for Scattered Disk Objects. Finally we interpreted the results under the assumption of hydrostatic equilibrium and found a lower limit for the density of 544$^{+42}_{-4}$ kg$\,$m$^{-3}$. However, a more likely density is 609$\pm$4 kg$\,$m$^{-3}$ using an aspect angle of 60$^\circ$, which corresponds to the most likely configuration for the spin axis with respect to the observer assuming random orientations.
\end{abstract}



\section{Introduction}

Trans-Neptunian Objects (TNOs) are bodies that orbit the Sun beyond the orbit of Neptune \citep[e.g.][]{Jewitt2008}. The first Trans-Neptunian Object to be discovered was 1992 QB$_1$ by \cite{Jewitt1993a}. Despite more than 20 years since this population was discovered, our overall knowledge about the physical properties of the objects that reside in the Trans-Neptunian Region is still scarce, mainly because of the faintness of these bodies. TNOs are thought to be mainly composed of mixtures of rocks and ice, a similar composition to that of comets \citep[e.g.][]{Barucci2011a, Elkins2010}. 

Due to their large distance to the Sun, TNOs are thought to be the least evolved in the Solar System. Hence they yield important information on the composition materials and physical conditions of the primitive solar nebula. The study of these bodies reveals plenty of information on the evolution of the Solar System since its initial phases. Additionally, the Trans-Neptunian Belt provides the natural connection with the study of protoplanetary disks observed around other stars.

Because of all the above we have been carrying out a rotational light curve survey of TNOs and Centaurs, for nearly two decades 
\citep[e.g.][]{Ortiz2002a,Ortiz2004,Ortiz2006a,Ortiz2007a,Ortiz2011a,Duffard2008,Santos-Sanz2008,Santos-Sanz2015,Lelouch2002a,Belskaya2006,Thirouin2010} to gather important physical information about them. In the course of our survey the TNO with provisional designation 2008 OG$_{19}$ showed interesting features such as a large amplitude of variability that encouraged us to make a very detailed study.

2008 OG$_{19}$ was discovered in July 2008 from Palomar Observatory and as far as we know the only published information on this body comes from a work by \cite{Sheppard2010a} in which data about several TNOs are reported. He obtained the absolute magnitude $(m_R(1,1,0)=4.47\pm0.02$ mag$)$ and the colours for this object \citep[see Table 2,][]{Sheppard2010a} based on the average of the photometry and assuming a phase slope parameter, but no further physical information on 2008 OG$_{19}$ was presented.

Here we present the first determination of the rotation period and the light curve amplitude for 2008 OG$_{19}$. The amplitude of the rotational light curve turned out to be remarkably high which was a clear indication of a very elongated shape for this body. Such elongated bodies are unusual within the Trans-Neptunian Belt, with Varuna being the archetype. We describe, in Section \ref{sec:observation}, the technical characteristics of the telescopes used for the observing runs. In Section \ref{data_reduction} and Section \ref{analysis} we describe the data reduction and the data analysis, respectively. The results from the light curve study are also included in Section \ref{analysis}. Furthermore, we report the method to obtain the absolute magnitude and we estimate the equivalent diameter of the target in Section \ref{absmag}. We discuss all the results in Section \ref{interpretation} and finally, we present our conclusions in Section \ref{conclusion}.



\section{Observations}
\label{sec:observation}

We took images of the TNO 2008 OG$_{19}$ in two observing runs during 2014 with different telescopes: the 1.23~m Calar Alto Observatory telescope in Almer\'ia (Spain) and the 1.5~m Sierra Nevada Observatory telescope in Granada (Spain).

The first observing run was on July 23, 29, 30 and 31 with the 4k$\times$4k DLR-MKIII CCD camera of the 1.23~m Calar Alto Observatory telescope. The image scale and the field of view (FOV) of the instrument are 0.32$^{\prime\prime}$pixel$^{-1}$ and 21\prp5$\times$21\prp5, respectively. The images were obtained in 2$\times$2 binning mode and were taken in the R-Johnson filter, the average seeing was 1\pri41 (see Table \ref{Observation}). We experienced good weather and dark nights (moonshine maximum $\sim$ 14\%). The signal to noise ratio (SNR) on the first night was $\sim$ 26 with 400 s exposure time, the SNR for the remaining ones was $\sim$ 40 with 300 s exposure time. The images of 2008 OG$_{19}$ were dithered over the detector to prevent problems in the photometry associated with bad pixels or CCD defects. Bias frames and twilight sky flat-field frames were taken each night to calibrate the images, and a total of 158 science images were obtained. We aimed the telescope at the same region of the sky each night to have the same stellar field throughout the observing run; that is important in order to choose the same comparison stars set for all nights in the observing run to minimize systematic photometric errors. This is possible thanks to the large enough FOV of the two telescopes that we have used.

\begin{table*}
	\centering
	\caption{Journal of Observation for 2008 OG$_{19}$. The July 23, 29, 30 and 31 days belong to the Calar Alto Observatory run. The August 21, 22, 24 and 25 days belong to the Sierra Nevada Observatory run.}
	\label{Observation}
	\begin{tabular}{cccccccccccc} 
		\hline
 UT Date (2014) & $J_D$ & Filters & Aper & $t_e$ & Seeing & R & $\Delta$ & $\alpha$ & $t_L$ & N & $t_{on-target}$\\
	
	 & & & (arcsec) & (seconds) & (arcsec) & (AU) & (AU)  & (deg) & (minutes) &  & (hours) \\
 \hline
 Jul 23 & 2456862.44292 &   R   & 2.56 & 400 & 1.23 & 38.5787 & 37.5806  & 0.290 & 312.5482 & 30 & 3.33\\
 Jul 29 & 2456868.46088 &   R   & 2.56 & 300 & 1.38 & 38.5788 & 37.5740  & 0.220 & 312.4926 & 33 & 2.75\\
 Jul 30 & 2456869.44575 &   R   & 3.84 & 300 & 1.61 & 38.5788 & 37.5738  & 0.217 & 312.4917 & 44 & 3.67\\
 Jul 31 & 2456870.41660 &   R   & 2.56 & 300 & 1.42 & 38.5789 & 37.5740  & 0.216 & 312.4933 & 51 & 4.25\\
 Aug 21 & 2456891.35487 & Clear & 3.71 & 400 & 1.53 & 38.5793 & 37.6440  & 0.574 & 313.0753 & 33 & 3.67\\
 Aug 22 & 2456892.35273 & Clear & 3.71 & 400 & 1.34 & 38.5794 & 37.6504  & 0.596 & 313.1285 & 37 & 4.11\\
 Aug 24 & 2456894.38882 & Clear & 3.71 & 400 & 1.40 & 38.5794 & 37.6640  & 0.640 & 313.2418 & 28 & 3.11\\
 Aug 25 & 2456895.33613 & Clear & 2.78 & 400 & 1.65 & 38.5794 & 37.6712  & 0.661 & 313.3017 & 44 & 4.89\\
		\hline
	\end{tabular}
	\begin{tablenotes}
      \small
      \item Filters based on the Johnson-Kron-Cousins system. Seeing calculated as the average for each night. Quantities are Universal Time date of the observation (UT Date), corresponding Julian Date for the first image of the night ($J_D$), aperture radii for photometry (Aper), exposure time ($t_e$), heliocentric distance (R), geocentric distance ($\Delta$), phase angle ($\alpha$), light travel time ($t_L$), number of images (N) and time on target each night ($t_{on-target}$).
    \end{tablenotes}
\end{table*}

The second observing run was on August 21, 22, 24 and 25 with the 2k$\times$2k CCD of the 1.5~m Sierra Nevada Observatory telescope. The image scale and the FOV of the instrument are 0.232$^{\prime\prime}$pixel$^{-1}$ and 7\prp92$\times$7\prp92, respectively. The images were obtained in 2$\times$2 binning mode and were taken with no filter to get the best SNR (which was $\sim$ 40 during the run), the exposure time was 400 s throughout the observing run. We had clear and dark nights; the moonshine was 17\% for the first night and 11\% for the second night, while the other two were completely dark nights. For this run the average seeing was 1\pri48, slightly worse than the first one. As in the July run from Calar Alto Observatory, we aimed again the telescope at the same coordinates all nights. Bias frames and twilight flat-field frames were taken each night to calibrate the exposures. In this run we took a total of 142 science images. In total, we analyzed 300 frames of 2008 OG$_{19}$.

All images were corrected for light travel time. A selection of the relevant observational data and 2008 OG$_{19}$'s orbital data are presented in Table \ref{Observation}.



\section{Data reduction}
\label{data_reduction}

To calibrate the images with bias frames and twilight sky flat-field frames we subtracted a median bias and divided by a median flat-field corresponding to each night. Specific routines written in IDL (Interactive Data Language) were developed for this task. The routines also included the code to perform the aperture photometry of all comparison stars and 2008 OG$_{19}$. We chose 22 stars for each run with good photometric behaviour. The aperture size was chosen in order to maximize the SNR on the Trans-Neptunian Object for each night and to minimize the dispersion of the photometry. We tried different apertures until the least dispersion in the residual to the fit to the experimental points in the photometry was obtained (see Section \ref{analysis}). This aperture corresponds to a radius between 4-6 pixels (2\pri56-3\pri84) and 6-8 pixels (2\pri78-3\pri71) for Sierra Nevada Observatory and Calar Alto Observatory, respectively (see Table \ref{Observation}). The median sky level was determined within an exterior annulus respect to the ring for the photometry with its inner radius equal to the size of the aperture plus 5 pixels, and a width of 5 pixels (in both telescopes). All images were checked in order to confirm the target was not close to any star (in that case the annulus used to subtract the sky background may be contaminated with light from other stars and the photometry might not be correct). At the time of observation the angular speed of 2008 OG$_{19}$ was 2.87$^{\prime\prime}$h$^{-1}$. The image trailing due to the motion of the object was only 0\pri23 and 0\pri31 during a typical 300 s and 400 s integration, respectively. This image trailing is negligible when we compare with our nominal 1\pri44 Full Width at High Maximum (FWHM) image quality. Through the aperture photometry we obtained the flux of all objects versus time (Julian Date). We obtained the relative photometry of the target with respect to each comparison star, so we had 22 light curves in total. To analyze them we calculated an average of all light curves and plotted this last one. The result are given in Table \ref{photometry}.

\begin{table}
	\centering
	\caption{Photometry results for the CAHA and OSN observations, respectively. In this table we list the Julian day (JD, corrected for the light time), the relative magnitude (Rel. Mag., in magnitudes), the error associated (Err., in magnitudes), the topocentric (r$_h$) and heliocentric ($\Delta$) distances (both distances expresed in AU) and the solar phase angle ($\alpha$, in degrees). The full table is aviable in online\label{photometry}.}
	\label{Fouriercoefficients}
	\begin{tabular}{cccccc} 
		\hline
 JD  & Rel. Mag.   & Err.  &   $\Delta$ & r$_\mathrm{h}$ & $\alpha$        \\
      & [mag]          &  [mag]&   [AU]      &  [AU]                   & [$^{\circ}$]    \\
 \hline
2456862.41613& 0.118&0.061&37.582&38.581&0.274\\
2456862.42088& 0.025&0.057&37.582&38.581&0.274\\
2456862.42565&-0.030&0.047&37.582&38.581&0.274\\
2456862.43041&-0.085&0.047&37.582&38.581&0.274\\
2456862.43517&-0.124&0.045&37.582&38.581&0.274\\
2456862.43993&-0.127&0.048&37.582&38.581&0.274\\
 
		\hline
	\end{tabular}
\end{table}



\section{Data Analysis and result}
\label{analysis}

\begin{figure*}
 \centering
  \subfloat[Lomb]{
   \label{periodogram}
    \includegraphics[angle=90,width=0.5\textwidth]{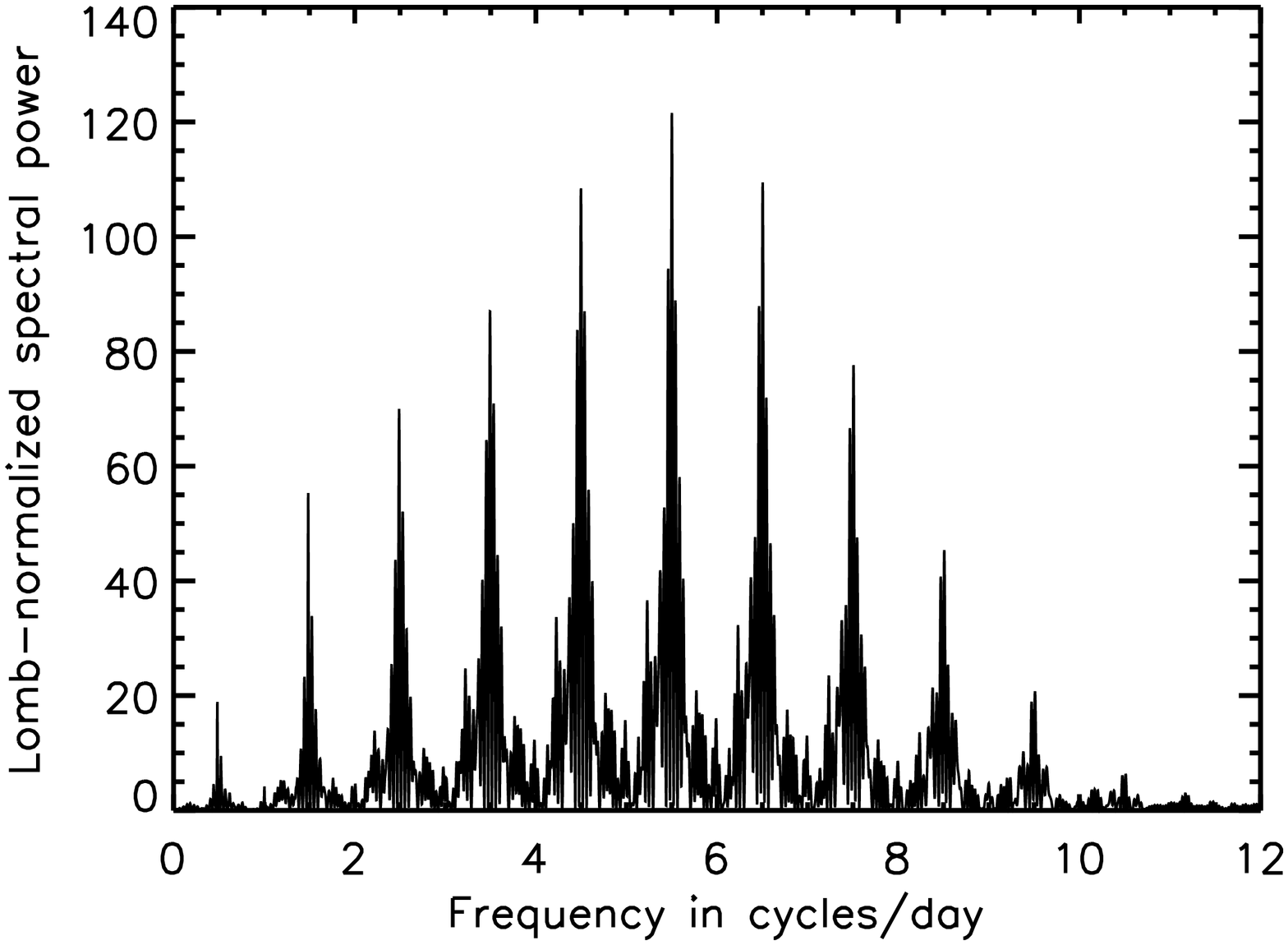}}
  \subfloat[PDM]{
   \label{PDM}
    \includegraphics[angle=90,width=0.5\textwidth]{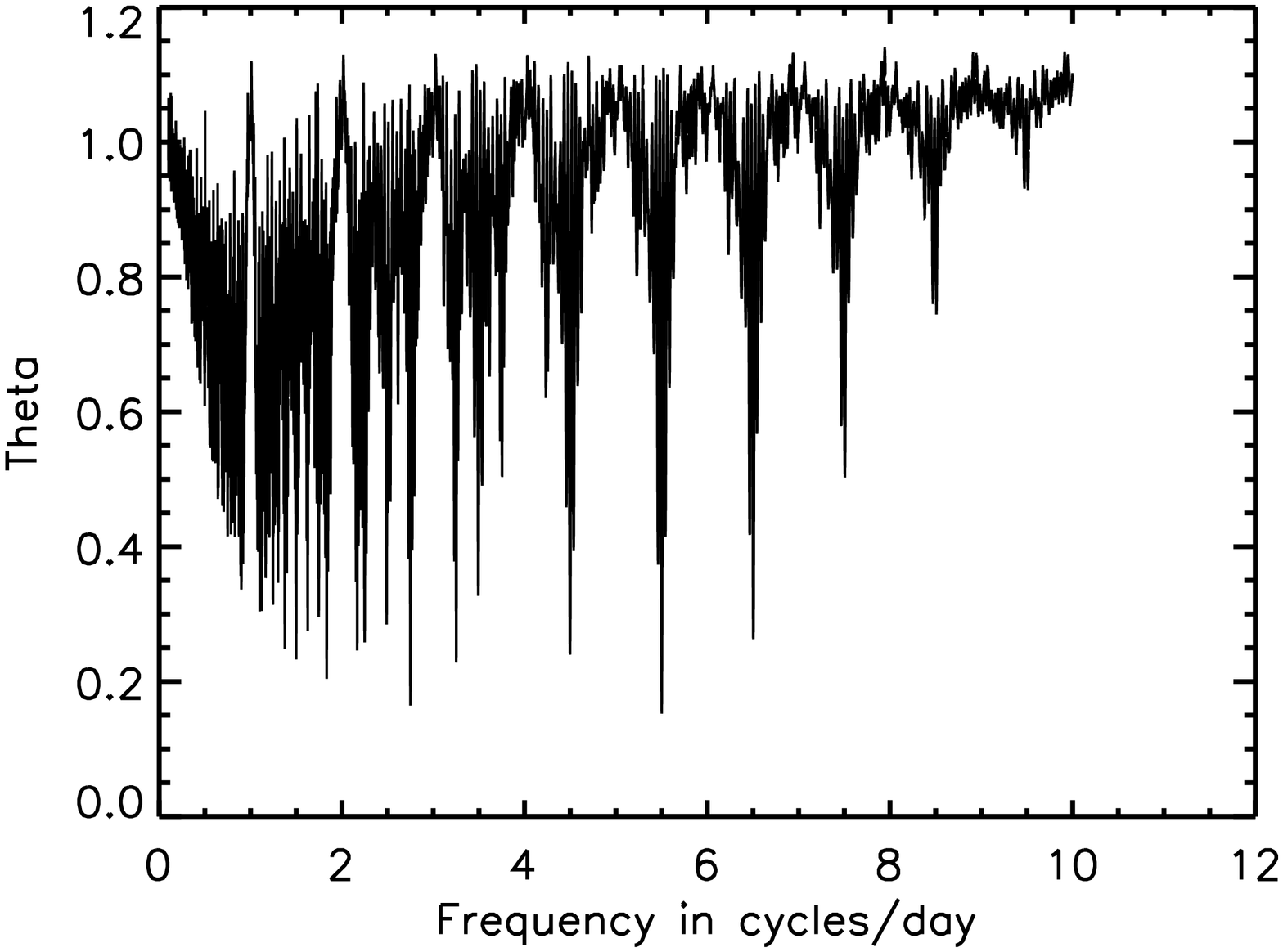}}
  \caption{(a) Lomb periodogram spectral power derived from the 2008 OG$_{19}$'s photometric data. The spectral power is plotted as a function of frequency (in cycles/day). The main peak is at 5.5 cycles/day (4.364 h) with a normalized spectral power of 121.6. The maximum for the spectral power is flanked by aliases due to the 24 h sampling periodicity. (b) Phase Dispersion Minimization plot computed from the photometric data. As can be seen, the minimum is at the frequency 5.50 cycles day$^{-1}$(4.364 h, which produces a single-peaked light curve). The second minimum is at the frequency 2.75 cycles day$^{-1}$ (8.727 h, which produces a double-peaked light curve).}
 \label{Lomb_PDM}
\end{figure*}

To obtain the rotational period of the target we applied two different techniques to the time series. The first one was a period search routine based on Lomb's technique \citep{Lomb1976} as implemented in \cite{Press1992}. This method is a modified version of the Fourier spectral analysis to take into account unevenly sampled data. The Lomb periodogram obtained is plotted in Fig. \ref{periodogram}; as can be seen the maximum spectral power peak was obtained for the frequency 5.5 cycles$\,$day$^{-1}$ (4.364 $\pm$ 0.001 h) with a normalized spectral power of 121.6. The maximum for the spectral power is flanked by aliases due to the 24-h sampling periodicity. The second technique was the Phase Dispersion Minimization (PDM) method. Contrary to the Lomb method, which searches the period that maximizes the normalized spectral power, PDM searches for the period that minimizes the so-called $\theta$ parameter \citep{Stellingwerf1978a}. Fig. \ref{PDM} shows the PDM $\theta$ parameter as a function of rotational frequency (we used frequency steps of 0.0005 cycles/day within the interval of 0.1 to 10 cycles/day). Two identical minima were obtained at 5.5 cycles$\,$day$^{-1}$ and 2.75 cycles$\,$day$^{-1}$ (4.364 $\pm$ 0.001 h and 8.727 $\pm$ 0.003 h, respectively). Therefore, we consider the 4.364 h and the 8.727 h periods as candidate rotational periods. 

\begin{figure}
	\includegraphics[width=\columnwidth]{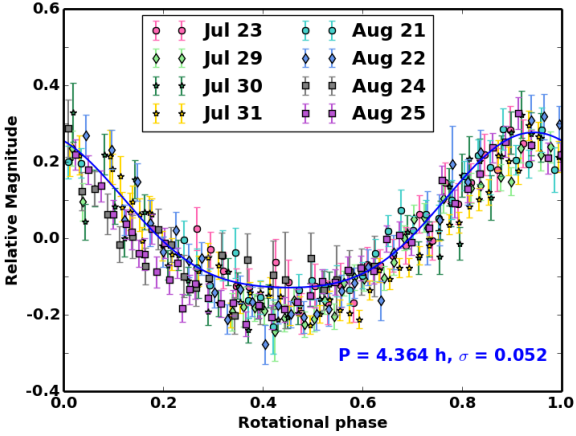}
    \caption{Folded light curve for the frequency corresponding to the maximum spectral power (P = 4.364 h). As can be seen, this is a single-peaked light curve. The blue line shows the fit of the Eq. \ref{equation_fit} to the points. $\sigma$ is the standard deviation of the residual to the fit.}
    \label{single_peaked}
\end{figure}

First, we folded the photometric data with the shorter candidate photometric period 4.364 h. This period produces a single-peaked light curve (Fig. \ref{single_peaked}). To calculate the single-peaked light curve amplitude, we fitted the data points using a Fourier function of second order, where $f_0$ is the rotational phase\footnote{Where we calculated the rotational phase as follow $f_0=(J_D-J_{D_0})/P$, where $J_D$ is the Julian day, $J_{D_0}=2456862$ is the initial Julian day, and P is the target's rotational period in days.}, as follows
\begin{equation}
a_0+a_1\cos{2\pi f_0}+b_1\sin{2\pi f_0}+a_2\cos{4\pi f_0}+b_2\sin{4\pi f_0},
\label{equation_fit}
\end{equation}in order to find the Fourier coefficients (a$_0$, a$_1$, a$_2$, b$_1$, b$_2$).

\begin{figure}
	\includegraphics[width=\columnwidth]{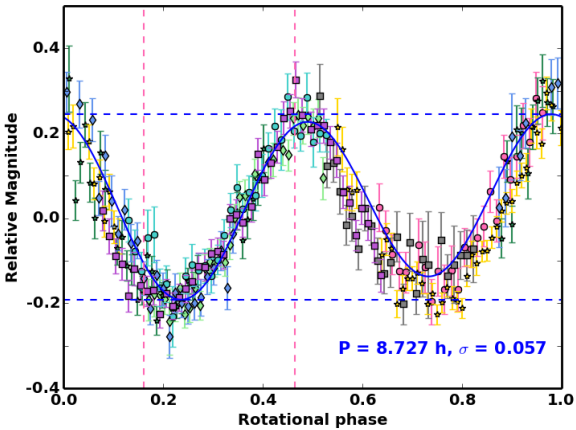}
    \caption{Folded light curve using the double of the period of the single-peaked light curve (P = 8.727 h). In this case, the produced light curve is double-peaked. The blue line shows the fit of the Eq. \ref{equation_fit} to the points. The dashed blue horizontal lines show the minimum and maximum of the fit and the dashed pink vertical lines show the rotational phases of Sheppard (2010) measurements. The legend for the data points is the same as in Fig. \ref{single_peaked}. $\sigma$ is the standard deviation of the residual to the fit.}
    \label{double_peaked}
\end{figure}

An amplitude ($\Delta m$) of $(0.406 \pm 0.011)$ mag was obtained for the single-peaked light curve. The folded data using the larger candidate period, P = (8.727 $\pm$ 0.003) h, show that the light curve of 2008 OG$_{19}$ has two maxima and two minima per rotation period (Fig. \ref{double_peaked}). The amplitude obtained from the fit to Eq. \ref{equation_fit} has, in this case, a value of $(0.437 \pm 0.011)$ mag. The Fourier coefficients for the fit of the double-peaked light curve (which is our preferred light curve, see section \ref{interpretation}) are given in Table \ref{Fouriercoefficients}. To estimate the period uncertainty we searched for all the periods that gave $\chi_{pdf}^2$ within $\chi^2_{pdf_{min}}$ and $\chi^2_{pdf_{min}}+1$. The $\chi_{pdf}^2$ value is derived from the fit to the Fourier function with the photometry data points, which was 2.63 for 8.727 h and 2.37 for 4.364 h.

\begin{figure}
	\includegraphics[width=\columnwidth]{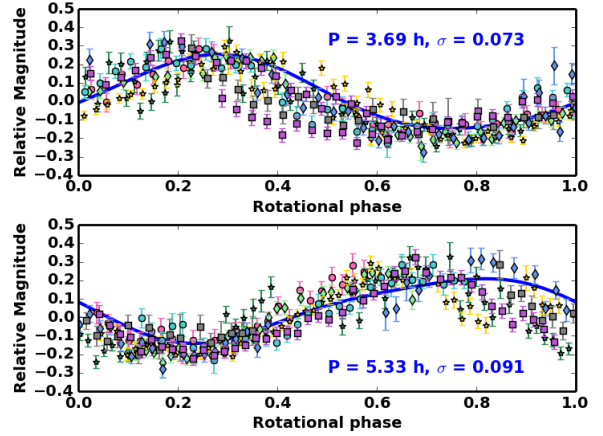}
    \caption{Folded light curves for the presumed 24 hr alias periods. Top panel: the light curve corresponds to a period of 3.69 h. Bottom panel: the light curve corresponds to a period of 5.33 h. The blue line, in both panels, shows the fit of the Eq. \ref{equation_fit} to the points. Visual inspection clearly indicate that those light curves are not acceptable. The legend for the data points is the same  as in Fig. \ref{single_peaked}. $\sigma$ is the standard deviation of the residual to the fit.}
    \label{ajuste_alias}
\end{figure}

The folded light curve produced using the 24 hr alias periods (P = 3.69 h, P = 5.33 h) are unconvincing and confirm that they are aliases (Fig. \ref{ajuste_alias}). We obtained the values of 3.96 and 6.03 for the $\chi_{pdf}^2$ test applied to fit to the Fourier function with the data points, for the periods of 3.69 h and 5.33 h, respectively. These values of $\chi_{pdf}^2$ are over the $\chi^2_{pdf_{min}}+1$ value that produced the double-peaked light curve period.

\begin{table}
	\centering
	\caption{Fourier coefficients of 2008 OG$_{19}$'s fitted light curve (P = 8.727 h).}
	\label{Fouriercoefficients}
	\begin{tabular}{ccccc} 
		\hline
	 n & $a_n$ & $\Delta(a_n)$ & $b_n$ & $\Delta(b_n)$ \\
 \hline
 0 & 0.036 & 0.004 &        &         \\
 1 & 0.006 & 0.005 & -0.028 & 0.005 \\
 2 & 0.196 & 0.005 & -0.039 & 0.005 \\
 
		\hline
	\end{tabular}
\end{table}



\section{Absolute Magnitude and equivalent diameter of 2008 OG$_{19}$}
\label{absmag}

An asteroid's absolute magnitude ($H$) is the visual magnitude when the asteroid is located at unit heliocentric and geocentric distances and at zero phase angle. The diameter of the asteroid ($D$) can be obtained from the absolute magnitude using the equation
\begin{equation}
D = C\,p^{-1/2}10^{-\frac{H}{5}},
\label{diameter}
\end{equation}where $C=1329$ km is a constant and both, $p$ and $H$, are the geometric albedo and the absolute magnitude of the object in the same photometric band, respectively \citep[e.g.][]{Russell1916b, Chesley2002}.

Because we did not observe Landolt reference stars in our Calar Alto and OSN runs, we had to calibrate the stellar fields that 2008 OG$_{19}$ traversed with auxiliary data. We took images of SA 112\_250 Landolt standard star \citep[which has similar colour as 2008 OG$_{19}$, according to the data by][]{Sheppard2010a} and the stellar field in a photometric night. We could not use neither the 1.23 m telescope at Calar Alto observatory or the 1.5 m telescope at Sierra Nevada Observatory because they were not available, so we observed with the IAC80 Telescope from Teide Observatory in Canary Islands (Spain) on November 24, 2014. The instrument used was CAMELOT CCD camera with 2048$\times$2048 pixels. The image scale and the FOV of the instrument are 0.304$^{\prime\prime}$pixel$^{-1}$ and 10\prp6$\times$10\prp6, respectively. The exposure time of the SA 112\_250 images was 10 s and the exposure time of the images of the same stellar region of Calar Alto run was 50 s.

We measured the flux of the Landolt standard star and the comparison stars and we obtained the comparison stars' magnitudes. Finally, with the comparison star's magnitudes we calculated 2008 OG$_{19}$'s absolute magnitude by means of the equation:
\begin{equation}
H_{R_i}=m_{star_i}-2.5\log\left(\frac{<F_{OG}>}{F_{star_i}}\right)-2.5\log\left({r_H}^n{\Delta}^2\right)-\phi(\alpha),
\label{absolutemag}
\end{equation}where $H_R$ is the R-band absolute magnitude, $m_{star_i}$ is the apparent magnitude of each comparison stars, $<F_{OG}>$ is an average of the flux during a rotational period of the target in ADUs, F$_{star_i}$ is the flux of each 22 comparison stars in ADUs, n is an index which reflects the change of the magnitude with the heliocentric distance (2 for asteroids), r$_H$ is the heliocentric distance in AU (Astronomical Units), $\Delta$ is the geocentric distance in AU, $\phi$ is the phase angle function and $\alpha$ is the phase angle in degrees. 

We calculated the phase angle correction by $\phi=\beta\alpha$, where $\beta= (0.30 \pm 0.12)$ mag deg$^{-1}$ was calculated by means of a linear regression via $m_R(1,1,\alpha)=H_R+\beta\alpha$ to two experimental points $(m_R(1,1,\alpha), \alpha)$. One given in \cite{Sheppard2010a} (4.66, 0.88), where he calculated the apparent magnitude based on the average of the photometry\footnote{The rotational phases of the photometric measures from \cite{Sheppard2010a} are plotted in Fig. \ref{double_peaked}, as can be seen, the two experimental points of Sheppard are near the minimum and the maximum on the light curve, hence, the average of the photometry match with the median level of the photometry} and the other experimental point was obtained in this study (4.48, 0.27), which is an average of the photometry. This $\beta$ value is slightly high for a TNO. Taking into account this $\beta$ value we obtained the absolute magnitude of 2008 OG$_{19}$ which is $(4.39 \pm 0.07)$ mag in R-band.

2008 OG$_{19}$'s orbital parameters are listed in Table \ref{comparison_2}. We found that 2008 OG$_{19}$ is a Scattered Disk Object following the DES classification \citep[its Tisserand parameter is equal to 3.03 greater than 3, and its eccentricity is greater than 0.2,][]{Elliot2005} and a Detached Object following the \cite{Gladman2008} classification (its eccentricity is higher than 0.24).

The mean geometric albedo in V-band (derived from Herschel Space Observatory data) for Scattered Disk and Detached Objects is 6.9\% and 17.0\%, respectively \citep[but note that the sample for Detached Objects is small]{Santos-Sanz2012}. We can estimate the geometric albedo in R band through:
\begin{equation}
\log\left(\frac{p_V}{p_R}\right)=\frac{2}{5}\left[(V-R)_{\odot}-(V-R)_{OG}\right],
\label{albedo}
\end{equation}where $p_V$ is the geometric albedo in V-band, $p_R$ is the geometric albedo in R-band, $(V-R)_{\odot}=0.36$ mag is the Sun's (V - R) and $(V-R)_{OG}=0.53$ mag is the 2008 OG$_{19}$'s (V - R) from \cite{Sheppard2010a}. Therefore, the geometric albedo in R-band is 8.1\% and 19.9\% for Scattered Disk Objects (SDOs) and Detached Objects (DOs), respectively.

Hence, an estimate of 2008 OG$_{19}$'s equivalent diameter is approximately $619^{+56}_{-113}$ km and $394^{+57}_{-63}$ km assuming albedos of 8.1\% and 19.9\% which are typical of SDOs and DOs. Varuna's equivalent diameter is $668^{+154}_{-86}$ \citep{Lellouch2013}, which is slightly bigger than the equivalent diameter obtained for 2008 OG$_{19}$ with a typical albedo of SDOs (the group in which Varuna is classified).



\section{Interpretation}
\label{interpretation}

For minor bodies in the Solar System a light curve is generally due to albedo variations on the surface, a non spherical body shape or a combination of both. In our particular case, we have two options, a single-peaked light curve and a double-peaked light curve. In the first case, this means that it would be due to an albedo variation on the surface. In the seconds case this means it would be due to a triaxial ellipsoid body shape. Therefore, whether the light curve is single-peaked or double-peaked has definite physical implications. In the following subsections we will provide several arguments pointing to the double-peaked interpretation as consistent with known physical models.

\subsection{Oblate Spheroid model}

TNOs suffer strong impacts that can convert them into a ``rubble pile" \citep{Farinella1981a}. After the impact, the ``rubble pile" will adopt a shape determined by the angular momentum and the density. Thus, this kind of objects can be treated as a fluid as first approximation. A fluid object with null angular momentum ($L=0$) is totally spherical. As L increases, the fluid deforms into an oblate spheroid, a Maclaurin Spheroid \citep{Chandrasekhar1987}. In our particular case, if 2008 OG$_{19}$'s rotational period is 4.364 h (the light curve is single-peaked with an amplitude of 0.405 mag), we obtained a lower limit for the density of 1699 kg$\,$m$^{-3}$. This density is unusually large for a typical TNO with the estimated equivalent diameter, 619 km if we assume that it belong to Scatter Disc Objects \citep{Ortiz2012b}.

Including some internal cohesion, not just fluid-like behavior we estimated the critical rotation period using \cite{Davidsson2001} formalism. For a typical density of TNOs \citep[$\approx$ 1000 kg$\,$m$^{-3}$, like Varuna's density,][]{Jewitt2002} and assuming an internal cohesion of 0.88$\rho$ \citep{Davidsson1999}, we obtained a critical rotation period of 4.837 h, slower than the rotation period that produce a single-peaked light curve. Hence, 2008 OG$_{19}$ would have broken up.

This leads us to think that it is very unlikely that 2008 OG$_{19}$ has an oblate shape and a rotation period of 4.364 h. Besides, rotational light curves caused by albedo features and capable of producing a very large light curve amplitude of 0.405 mag are not known in the Kuiper Belt \citep[e.g.][]{Jewitt2002, Duffard2009, Gutierrez2001, Ortiz2003a}. Therefore, the natural interpretation of the rotational light curve of 2008 OG$_{19}$ is that it results from the cyclic variations of the cross section of a triaxial ellipsoid as seen from an observer on Earth.

\subsection{Triaxial ellipsoid model}

We considered that the 2008 OG$_{19}$'s rotational period is 8.727 h. Hence, the body produces a double-peaked light curve due to a prolate ellipsoid body shape, with an amplitude of 0.437 mag. As we stated above, a fluid object with null angular momentum is totally spherical, as L increases, the fluid deforms into an oblate spheroid. This happens until a critical value $L$ = 0.304 from which the body becomes a triaxial Jacobi ellipsoid with semi major/minor axis $a$/$b$, $c$ (where $a>b>c$) \citep{Chandrasekhar1987}. The axis ratio can be obtained from the known equation
\begin{equation}
\Delta\,m=2.5\log{\left(\frac{a}{b}\right)},
\label{axis_ratio}
\end{equation} where $\Delta{m}$ is the double-peaked light curve amplitude. We obtained an axis ratio $a/b\approx 1.495$. Hence, we estimated a density $\rho\,\ge$ 544 kg$\,$m$^{-3}$ and an angular momentum $L$ = 0.324  (from Table IV of \cite{Chandrasekhar1987}), higher than 0.304. This result is valid if we think that the 2008 $OG_{19}$'s aspect angle (the angle between the rotation axis and the line of sight) is 90$^\circ$. Nevertheless, in a random distribution of spin vectors, the probability of observing an object with an aspect angle in the range [$\delta$,($\delta+d\delta$)] is proportional to $\sin(\delta)d\delta$. The average aspect angle $\delta$ is thus 60$^{\circ}$ \citep{Sheppard2004}. Therefore, if one supposes that the most probable aspect angle for TNOs is 60$^\circ$, the light curve amplitude of a triaxial body with an arbitrary aspect angle is given by the equation
\begin{equation}
\Delta\,m=-2.5\log\left[{\frac{b}{a}\left(\frac{a^2\cos^2(\delta)+c^2\sin^2(\delta)}{b^2\cos^2(\delta)+c^2\sin^2(\delta)}\right)^{\frac{1}{2}}}\right],
\end{equation}where $\delta$ is the aspect angle (as can be seen, for $\delta=90^\circ$ it leads to Eq. \ref{axis_ratio}). Using an aspect angle of 60$^\circ$, we obtained an axis ratio $b/a$ of 0.513, this implies an axis ratio $c/a=0.390$ and a density of 609 kg$\,$m$^{-3}$. 

On the other hand, as can be seen in Fig. \ref{double_peaked}, the maxima and the minima of the light curve have different depths, which is another indication that the light curve is indeed due to the body shape, because real bodies with perfect symmetry are very unlikely.

Thus 609 kg$\,$m$^{-3}$ is the more likely density of 2008 OG$_{19}$ and it is a reasonable value according to the density versus size plot in the supplementary material of \cite{Ortiz2012b}.

Taking into account the light curve and the light curve amplitude, Varuna is a similar object to 2008 OG19. However, in terms of the derived density, 2008 OG$_{19}$ has a smaller density than that of Varuna. If we assume that the albedo of 2008 OG$_{19}$ is similar to that of Varuna, then 2008 OG$_{19}$ would be smaller than Varuna and the smaller density of 2008 OG$_{19}$ would make sense in this context, because it is already known that small TNOs are less dense than larger TNOs \cite[see Fig. S7 in the supplementary material of ][]{Ortiz2012b}. The most likely explanation for this is a larger porosity in smaller objects, whereas for large TNOs one can expect that the material is more compacted.

Varuna has a larger albedo than the average albedo of SDOs, possibly resulting from the water ice spectroscopically detected on its surface but unfortunately the albedo of 2008OG19 is not known and spectra of 2008 OG$_{19}$ are also lacking. Any spectral evidence for water ice like in Varuna, might be consistent with a larger than average geometric albedo. Regarding V-R color, Varuna and 2008OG19 are similar but no information on the albedo can be derived from that. A comparison table of Varuna versus 2008 OG$_{19}$ in terms of physical properties is shown in Table \ref{comparison_1}. Regarding orbital parameters, 2008 OG$_{19}$ is more similar to Eris except for the inclination, which is higher for Eris than for 2008 OG19's case (Table 4).

\begin{table*}
\begin{threeparttable}[b]
	\caption{Physical properties of both: 2008 OG$_{19}$ and Varuna.}
	\label{comparison_1}
	\begin{tabular}{ccccccccc} 
		\hline
	 TNO         & $\Delta m$    & p$_V$                         & $V-R$        & $H_R$        & P               & D                            & $\rho$        & $\beta$ \\
	             & (mag)         & (\%)                          & (mag)        & (mag)        & (h)             & (km)                         & (kgm$^{-3}$)  & (mag deg$^{-1}$) \\
 \hline
 2008 OG$_{19}$& 0.437\tnote{1}& 6.9 (SDOs)/17.0 (DOs)\tnote{4}& 0.53\tnote{5}& 4.39\tnote{1}& 8.727\tnote{1}  & 619 (SDOs)/394 (DOs)\tnote{1}& 609\tnote{1}  & 0.30\tnote{1} \\
\hline
 Varuna        & 0.44\tnote{2} & 12.7\tnote{3}                 & 0.64\tnote{6}& 2.95\tnote{2} & 6.34358\tnote{2}& 668\tnote{3}                 & 1000\tnote{2} & 0.10\tnote{2}\\
               & 0.42\tnote{6} &                               &              & 3.192\tnote{7}&                 &                              &               & 0.156\tnote{6}\\
               & 0.43\tnote{8} &                               &              &               &                 &                              &               &\\
		\hline
	\end{tabular}
	\begin{tablenotes}
	\item [1] This work;
	\item [2] \cite{Belskaya2006};
	\item [3] \cite{Lellouch2013};
	\item [4] assumed from \cite{Santos-Sanz2012};
	\item [5] \cite{Sheppard2010a};
	\item [6] \cite{Jewitt2002};
	\item [7] \cite{Hicks2005};
	\item [8] \cite{Thirouin2010}.\\
	\item Quantities are light curve amplitude ($\Delta m$), albedo in V-band (p$_V$), $(V-R)$ Colour, absolute magnitude in R-band ($H_R$), period (P), equivalent diameter (D), density ($\rho$) and phase slope parameter ($\beta$).
	\end{tablenotes}
	\end{threeparttable}
\end{table*}

\begin{table}
\begin{threeparttable}[b]
	\caption{Comparison of orbital parameters of 2008 OG$_{19}$, Varuna and Eris.}
	\label{comparison_2}
	\begin{tabular}{cccccc} 
		\hline
	 TNO          & e & a & q & i \\
	              &  & (AU) & (AU) & (deg) \\
 \hline
 2008 OG$_{19}$ & 0.417 & 66.197 & 38.577 & 13.165\\
 Varuna         & 0.051 & 43.161 & 40.972 & 17.156\\
 Eris           & 0.441 & 67.729 & 37.825 & 44.099\\
 
		\hline
	\end{tabular}
		\begin{tablenotes}
	\item Quantities are eccentricity (e), semi-major axis (a), perihelion distance (q) and inclination (i). Data taken from Minor Planet Center.
	\end{tablenotes}
	\end{threeparttable}
\end{table}

\subsection{Binary Body model}

We can also consider the possibility that 2008 OG$_{19}$ is a binary object, in which case the light curve could be due to the occultation of one component by the other. Using the same methodology as in \cite{Jewitt2002}, we obtained a lower limit for the density of 514 kg$\,$m$^{-3}$, very close to the density that we obtained from triaxial Jacobi ellipsoid model with an aspect angle of 90$^\circ$, so it is possible that the binary model can account for 2008 OG$_{19}$ (like the case of Varuna). But we don't have any evidence that 2008 OG$_{19}$ is a binary object and usually the eclipses produce minima and maxima with less soft curvatures \citep[e.g.][]{Noll2003}.



\section{Conclusions}
\label{conclusion}

We think that the light curve of 2008 OG$_{19}$ is caused by the rotation of a body of triaxial Jacobi shape, similar to that of Varuna. Both objects have a double-peaked light curve with diferent depths for the two minima and two maxima. 2008 OG$_{19}$'s peak to valley amplitude is $(0.437 \pm 0.011)$ mag (close to Varuna's amplitude) with a rotational period of (8.727 $\pm$ 0.003) h. We also found that 2008 OG$_{19}$'s absolute magnitude in R-band is $(4.39 \pm 0.07)$ mag, with $\beta$ $\approx$ 0.30 mag deg$^{-1}$, and that 2008 OG$_{19}$'s equivalent diameter is 619 km using a typical albedo for Scattered Disc Objects (8.1\%) and 394 km using a typical albedo for Detached Objects (19.9\%). The preferred density for 2008 OG$_{19}$ is 609 kg$\,$m$^{-3}$ assuming hydrostatic equilibrium and a 60$^{\circ}$ aspect angle.


\section*{Acknowledgements}
%
We are grateful to the CAHA and OSN staffs. This research is partially based on observations collected at Centro Astron\'omico Hispano Alem\'an (CAHA) at Calar Alto, operated jointly by the Max-Planck Institut fur Astronomie and the Instituto de Astrof\'isica de Andaluc\'ia (CSIC). This research was also partially based on observation carried out at the Observatorio de Sierra Nevada (OSN) operated by Instituto de Astrof\'isica de Andaluc\'ia (CSIC). This article is also based on observations made with the telescope IAC80 operated on the island of Tenerife by the Instituto de Astrof\'isica de Canarias in the Spanish Observatorio del Teide. Funding from Spanish grant AYA-2014-56637-C2-1-P is acknowledged, as is the Proyecto de Excelencia de la Junta de Andaluc\'ia, J. A. 2012-FQM1776. R.D. acknowledges the support of MINECO for his Ramon y Cajal Contract. FEDER funds are also acknowledged. We thank an anonymous referee for helpful comments.






%
%


\bsp	
\label{lastpage}
\end{document}